\documentclass{jicspack}
\setvolume{8}                             
\setyear{2012}                             
\setpagerange{1}{8}                    
\setheadauthor{X. Liu et al.}          
\setissn{1553--9105} \setpubdate{January 2012} \setno{1}

\afterpage{\beginheader}                   

\usepackage{enumerate}
\usepackage{graphicx}

\usepackage{amssymb}

\begin{document}

\begin{premaker}


\title{What is   Cook's theorem?}
\author{JianMing ZHOU, Yu LI}
\ead{yu.li@u-picardie.fr}
\address{(1) MIS, Universit\'{e} de Picardie Jules Verne, 33 rue Saint-Leu, 80090 Amiens, France \\
(2) Institut of computational theory and application, Huazhong University of Science and Technology, Wuhan, China}


\begin{abstract}

 In this paper, we  make a preliminary interpretation of Cook's theorem presented  in \cite{cook1}. This interpretation reveals   cognitive biases in the proof of Cook's theorem that arise from the attempt of  constructing a formula in $CNF$ to represent a computation of a nondeterministic Turing machine. Such cognitive biases are  due to  the lack of understanding about the essence of  \textit{nondeterminism}, and lead to  the   confusion between different  levels of  \textit{nondeterminism} and  \textit{determinism}, thus  cause the loss of \textit{nondeterminism}  from the $NP$-completeness theory.  The work shows that Cook's theorem is  the origin of the loss of \textit{nondeterminism}  in terms of   the equivalence of  the two definitions of $NP$,  the one  defining $NP$ as the class of problems solvable by a nondeterministic Turing machine in polynomial time, and the other  defining $NP$ as the class of problems verifiable by a deterministic Turing machine in polynomial time. Therefore, we argue that   fundamental difficulties in understanding \textit{P versus NP}  lie firstly  at cognition level, then logic level.

\end{abstract}
\begin{keyword}
Cook's theorem \sep $CNF$  \sep $P$ versus $NP$ \sep  $NDTM$ (NonDeterministic Turing Machine) \sep $DTM$ (Deterministic Turing Machine) \sep  oracle  \sep query machine \sep  $NDTM$ model   \sep equivalence of the two definitions of $NP$ 
\end{keyword}

\end{premaker}

\section{Introduction}

The notion of \textit{nondeterminism} is lost from the current definition of $NP$, which is reflected  in the equivalence of the two definitions of $NP$   commonly accepted   in the academic community   \cite{np}\cite{sipser}\cite{cook2}\cite{garey},  the one is the solvability-based definition that  defines $NP$ as   the class of problems solvable  by a nondeterministic Turing machine in polynomial time,  and the other is the verifiability-based definition that defines $NP$ as the class of problems verifiable by a deterministic Turing machine in polynomial time.  Due to this equivalence, the verifiability-based definition  has been accepted as the standard definition of $NP$, which has led to the disappearance of  nondeterminism   from $NP$, and caused  ambiguities in understanding $NP$, thus  \textit{P versus NP}  \cite{william}. 
 
In  the paper entitled \textit{What is $NP$? - Interpretation of  a Chinese  paradox: White horse is not horse} \cite{li2},   we  questioned this equivalence.   With the help of a famous Chinese paradox \textit{White horse is not horse}, we  interpreted some well-known  arguments supporting this equivalence,  and revealed cognitive biases   that  cause  the confusion between different levels of    \textit{nondeterminism} and  \textit{determinism} from the view of recognition of problem.   

 In this paper,  we make a preliminary interpretation of Cook's theorem presented  in \cite{cook1}  from the view of representation of  problem, and reveal cognitive biases  in  the proof of Cook's theorem.

The paper is organized as follows.  In Section 2, we present an overview of Cook's theorem.  In Section 3, we interpret  Cook's theorem based on  \textit{query machine}.  In Section 4, we interpret   Cook's theorem based on   \textit{NDTM model}. In Section 5, we propose to rectify Cook's theorem,  and  in Section 6 we conclude the paper.
 
 \section{Overview of Cook's theorem}
  
 Cook's theorem is usually stated as \cite{cook3}:
 
\textit{Any problem in $NP$ can be reduced in polynomial time by a deterministic Turing machine to the problem of determining whether a formula in $CNF$ is satisfiable  ($SAT$).} \\ 
    
However, the original statement of Cook's theorem was presented  in Cook's paper entitled   \textit{The complexity of theorem proving procedures}  as \cite{cook1}:

\textbf{Theorem 1}
\textit{If a set $S$ of strings is accepted by some nondeterministic Turing machine within polynomial time, then $S$ is $P$-reducible to \{$DNF$ tautologies\}}.  \\

The main idea of the proof of  \textbf{Theorem 1}  was described  in \cite{cook1}:

 \textit{Suppose a nondeterministic Turing machine $M$ accepts a set $S$ of strings within time $Q(n)$, where $Q(n)$ is a polynomial. Given an input $w$ for $M$, we will construct a propositional formula $A(w)$ in conjunctive normal form ($CNF$) such that $A(w)$ is satisfiable iff $M$ accepts $w$.  Thus $\neg A(w)$ is easily put in disjunctive normal form (using De MorganÕs laws), and $\neg A(w)$ is a tautology if and only if w $ \not \in S$. Since the whole construction can be carried out in time bounded by a polynomial in $\mid w \mid$ (the length of $w$), the theorem will be proved.}\\
 
Here  $S$ refers to  a set of all instances   of an  $NP$ problem that have solutions,  and   finding the  tautology of  $\neg A(w)$ in $DNF$ is transformed into   finding the  satisfiability of  $A(w)$ in $CNF$. 

Concerning  \textit{$P$-reducibility}, it  was explained in  \cite{cook1}:
 
  \textit{Here  "reduced" means, roughly specking, that the first problem can be solved deterministically in polynomial time provided an oracle is available for solving the second. }\\

That is, Cook attempted to   construct a formula $A(w)$ in $CNF$ to represent a \textit{computation of a nondeterministic Turing machine}  in order to achieve the objective of representing an $NP$ problem    as  $SAT$ problem, however the construction of $A(w)$  is a deterministic and polynomial time process, that is,  a \textit{computation of a  deterministic Turing machine}. Therefore, how to construct  $A(w)$ constitutes the proof of  \textbf{Theorem 1}.  

We interpret this proof and  reveal  cognitive biases hidden in it. 
 
  \section{Interpretation of Cook's theorem based on  \textit{Query Machine}}
        
 \subsection{Query Machine and P-reduciblility}  

In order to provide a reasonable basis for \textit{P-reduciblility}  in  \textbf{Theorem 1},  Cook  introduced a tool   called \textit{query machine}, which  is a mix of an oracle and a deterministic Turing machine, to replace a nondeterministic Turing machine. It is this \textit{query machine} that sows   the seeds of    confusion of  $NDTM$ (NonDeterministic Turing Machine) and  $DTM$ (Deterministic Turing Machine) in the proof of  \textbf{Theorem 1}.\\
  
Let us analyze this query machine.  A  query machine  was  defined   in  \cite{cook1}:

  \textit{A query machine is a multitape Turing machine with a distinguished tape called the query tape, and three distinguished states called the $query~state$, $yes~state$, and $no~state$, respectively. If $M$ is a query machine and $T$ is a set of strings, then a $T$-computation of $M$ is a computation of $M$ in which initially $M$ is in the initial state and has an input string $w$ on its input tape,  and each time $M$ assures the query state there is a string $u$ on the query tape,  and the next state $M$ assumes is the yes state if $u \in  T$ and the no state if $u  \not  \in T$.  We think of an 'oracle', which knows $T$, placing $M$ in the yes state or no state}.  \\

  Then the concept of  \textit{P-reduciblility} was defined based on  query machine \cite{cook1}:
  
  \textit{Definition. A set S of strings is P-reducible (P for polynomial) to a set T of strings iff there is some query machine M and a polynomial Q(n) such that for each input string  w, the T-computation of M with input w halts within Q($\mid w \mid$) steps ($\mid w \mid$ is the length of w) and ends in an accepting state iff  $w    \in S$}. 

\textit{It is not hard to see that P-reducibility is a transitive relation. Thus the relation $E$ on sets of strings, given by $(S,T)  \in E$ iff each of $S$ and T is P-reducible to the other, is an equivalence relation.  The equivalence class containing a set S will be denoted by deg (S) (the polynomial degree of difficulty of S).}\\

 In addition,  five $NP$ problems were given as examples  to illustrate $S$ and $T$ \cite{cook1}:
 
 \textit{We now define the following special sets of strings.}
 \begin{enumerate}

\item \textit{The subgraph problem is the problem given two finite undirected graphs, determine whether the first is isomorphic to a subgraph of the second. A graph G can be represented by a string G on the alphabet $ \{0,1,*\}$ by listing the successive rows of its adjacency matrix, separated by $*$s. We let {subgraph pairs} denote the set of strings $\overline G_1\ast\ast \overline G_2$ such that $G_1$ is isomorphic to a subgraph of $G_2$.}

\item \textit{The graph isomorphism problem will be represented by the set, denoted by $\{$isomorphic graph pairs$\}$, of all strings $\overline G_1\ast\ast \overline G_2$ such that $G_1$ is isomorphic to $G_2$.}

\item \textit{The set $\{$Primes$\}$ is the set of all binary notations for prime numbers.}

\item \textit{The set $\{$DNF tautologies$\}$ is the set of strings representing tautologies in disjunctive normal form.}

\item \textit{The set D3 consists of those tautologies in disjunctive normal form in which each disjunct has at most three conjuncts (each of which is an atom or negation of an atom).}\\

 \end{enumerate}

 \begin{figure} [h]
\begin{center}
\includegraphics[scale=.7]{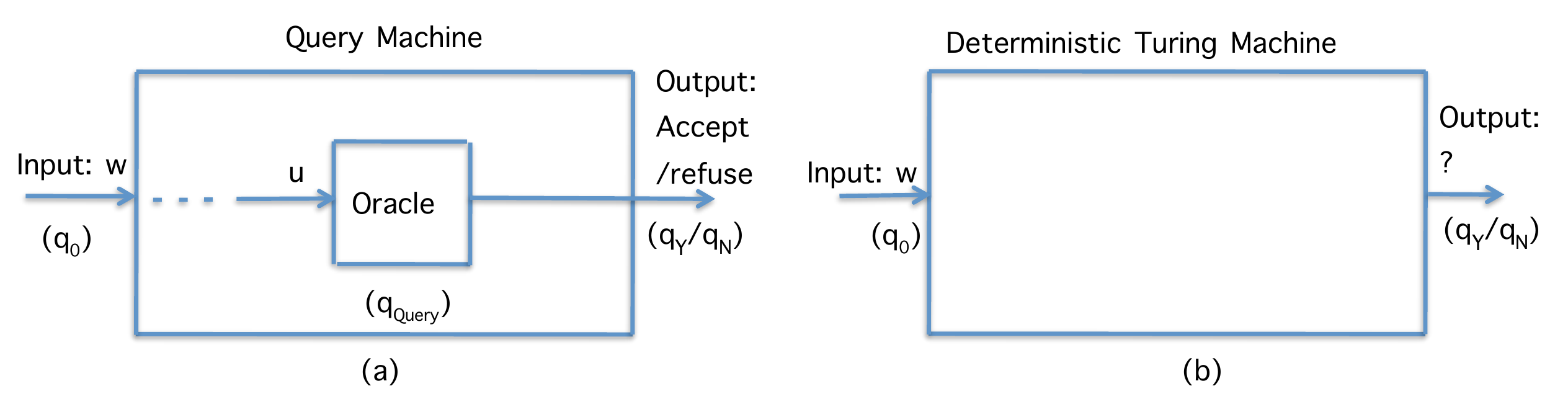}
\end{center}
\caption{(a) A query machine accepts $w$; (b) A deterministic Turing machine accepts $w$}
\label{fig5}
\end{figure}

We  interpret how a query machine $M$ accepts an instance $w$ of an $NP$ problem in polynomial time, which is illustrated in Fig. 1(a).

$S$ refers to a set of strings that represents all instances that have solutions, for example,  $S$ refers to a set of instances $\overline G_1 \ast\ast \overline G_2$ of the  graph isomorphism problem  such that $G_1$ is isomorphic to $G_2$.  $T$ refers to a set  of formulas in $DNF$ that are tautologies.

Initially, $M$ is in the initial state $q_0$ and has  $w$   representing an instance of a $NP$ problem as input. Then, $M$   assures the query state $q_{Query}$ where there is a string $u$ representing a formula in $DNF$ as input for an oracle and this oracle instantly determines whether $u \in  T$, that is, whether $u$ is  tautology. Finally, according to the  obtained reply,   if $u \in  T$ then  the oracle places $M$ in the yes state $q_Y$ and accepts $w$;   or if $u  \not  \in T$ then   the oracle places $M$ in the no state $q_N$ and refuses $w$.   

In this way, $S$ is said to be $P$-reducible to $T$.\\

Therefore, a query machine is in fact a formalized \textit{oracle}. But it should pay special attention to the essence of oracle. The existence of an oracle is just a hypothesis rather than a fact, so it is just theoretically valid, but not in practice, while a deterministic Turing machine is either  theoretically or  practically  valid. That is, \textit{oracle} and $DTM$  are two concepts situated at  different levels  (see Fig. 1).

Unfortunately, it seems that Cook did not realize this fundamental difference, when he interpreted  $P$-reducibility in  \cite{cook1}: 

 \textit{By reduced we mean, roughly speaking, that if tautology hood could be decided instantly (by an "oracle") then these problems could be decided in polynomial time. In order to make this notion precise, we introduce query machines, which are like Turing machines with oracles in [1].} \\

In other words,  Cook  made a direct logic deduction between \textit{oracle} and $DTM$, but this is a  \textit{question begging}, because the existence of an oracle itself needs to be proved. It is  this cognitive bias that   leads to the confusion of \textit{query machine} and $DTM$, thus  the confusion of  $NDTM$ and $DTM$  in the proof of \textbf{Theorem 1}.

 \subsection{Proof of \textbf{Theorem 1}}
 
The   proof of \textbf{Theorem 1} consists in   constructing $u$ from $w$, where  $w$ is an instance of an $NP$ problem and $u$ is a  formula  $A(w)$ in $CNF$  (see   Fig. 1(a)),  through representing a computation of a  nondeterministic Turing machine  for $w$. 

This idea  was explained in    \cite{cook1}:
 
 \textit{Suppose a nondeterministic Turing machine $M$ accepts a set $S$ of strings within time $Q(n)$, where $Q(n)$ is a polynomial. Given an input $w$ for $M$, we will construct a propositional formula $A(w)$ in conjunctive normal form ($CNF$) such that $A(w)$ is satisfiable iff $M$ accepts $w$.} \\
 
However,  the construction of $A(w)$ is a deterministic and polynomial time process, so Cook discretely replaced  a nondeterministic Turing machine by a deterministic Turing machine through a  query machine  (see    Fig. 1).  

We interpret how it could happen.  \\

Originally the above $M$  refers to a nondeterministic Turing machine, but this $M$ has  discretely  changed   in the following \cite{cook1}:

 \textit{We may as well assume the Turing machine has only on tape, which is infinite to the right but has a left-most square. Let us number the squares from left to right $1,2,\dotsc$ Let us fix an input $w$ to $M$ of length $n$, and suppose $w \in S$. Then there is a computation of M with input w that ends in an accepting state within $T=Q(n)$ steps. The formula $A(w)$ will be built from many different proposition symbols, whose intended meaning, listed below, refer to such a computation.}\\

That is, $A(w)$ is built to represent a computation of $M$. Let us interpret the meaning of   this computation  through the analysis of  the  construction of $A(w)$  \cite{cook1}:

 \textit{Suppose the tape alphabet for $M$ is $ \{ \sigma_1,\dotsc, \sigma_l \}$  and the set of states is $ \{q_1,\dotsc, q_r \}$. Notice that since the computation has at most $T=Q(n)$ steps, no tape square beyond $T$ is scanned.}

 \textit{Proposition symbols :}
 \textit{\begin{itemize}
\item $P_{s,t}^i$  for $1\leq i \leq l$, $1\leq s,t \leq T$. $P_{s,t}^i$  is true iff tape square number $s$ at step $t$ contains the symbol $\sigma_i$. 
\item  $Q_{t}^i$ for $1 \leq i \leq r$, $1\leq t \leq T $. $Q_{t}^i$ is true iff at step t the machine is in state $q_i$.
\item  $S_{s,t}$ for $1\leq s,t \leq T)$ is true iff at step $t$ square number $s$ is scanned by the tape head.
\end{itemize}
}

 \textit{The formula $A(w)$ is a conjunction $B  \land C   \land  D   \land  E  \land  F   \land  G  \land  H   \land  I$ formed as follows. Notice A(w) is in conjunctive normal form. }

 \textit{B will assert that at each step t, one and only one square is scanned. B is a conjunctive ${B_1   \land  B_2   \land  \dotsc  \land  B_{T}}$, where $B_t$ asserts that at time t one and only one square is scanned :}

 \textit{$B_t = (S_{1,t}  \lor  S_{2,t} \lor ,\dotsc \lor  S_{T,t})    \land  \big [ \bigwedge_{1\leq i<j \leq Q(n)} ( \neg S_{i,t} \lor  \neg S_{j,t}) \big].$}

 \textit{For $1\leq s \leq T$ and $1\leq t \leq T$, $C_{s,t}$ asserts that at square $s$ and time $t$ there is one and only one symbol. $C$ is the conjunction of all the $C_{s,t}$.}

 \textit{$D$ asserts that for each $t$ there is one and only one state.}
 

 \textit{$E$ asserts the initial conditions are satisfied :}

 \textit{$E = Q_1^0   \land   S_{1,1}   \land   P_{1,1}^{i_1}  \land   P_{2,1}^{i_2}  \land \dotsc   \land  P_{n,1}^{i_n}  \land  P_{n+1,1}^1 \dotsc  \land  P_{T,1}^1$}

 \textit{Where $w =  \sigma_{i_1}, \dotsc, \sigma_{i_n}$   , $q_0$ is the initial state and $\sigma_1$  is the blank symbol.}


 \textit{$F$, $G$, and $H$ assert that for each time $t$ the values of the $P'$s, $Q'$s and $S'$s are updated properly. For example, $G$ is the conjunction over all $t$, $i$, $j$ of $G_{i,j}^t$, where $G_{i,j}^t$  asserts that if at time $t$ the machine is in state $q_i$ scanning symbol $\sigma_{j}$, then at time $t+1$ the machine is in the state $q_k$, where $q_k$ is the state given by the transition function for $M$.}

 \textit{$G_{i,j}^t  =  \bigwedge_{s=1}^{T} ( \neg Q_t^i   \lor  \neg S_{s,t}   \lor  \neg P_{s,t}^j   \lor Q_{t+1}^k).$}
 

 \textit{Finally, the formula $I$ asserts that the machine reaches an accepting state at some time. The machine $M$ should be modified so that it continues to compute in some trivial fashion after reaching an accepting state, so that $A(w)$ will be satisfied.}\\

Firstly,  let us  clarify  $M$. Notice that $B$, $C$, $D$  and $I$ refer to just the physical structure of  $M$;  $F$, $G$ and $H$ refer to  the transition function of $M$; and $E$  concerns the initial condition of a computation of $M$. 

Such a computation  proceeds as below. Starting from $E$, the assignment of the proposition symbols corresponding to following steps are deduced out according to $B  \land C   \land  D    \land  F   \land  G  \land  H$,  finally  the truth of  $I$ is deduced out in polynomial time.  Since this computation  ends in polynomial time $Q(n)$,  and $F$, $G$ and $H$ represent  the transition function of $M$ in terms of  \textit{$G_{i,j}^t  =  \bigwedge_{s=1}^{T} ( \neg Q_t^i   \lor  \neg S_{s,t}   \lor  \neg P_{s,t}^j   \lor Q_{t+1}^k)$}, then  this computation is in fact a computation of a deterministic Turing machine $M$. In other words, the original nondeterministic Turing machine $M$ has been discretely  transformed into a deterministic Turing machine $M$! \\

Then, we  interpret the meaning of such a computation by clarifying  $E$. $E$  refers to  the assignment of the proposition symbols corresponding to the initial time $t=1$,   \textit{$E = Q_1^0   \land   S_{1,1}   \land   P_{1,1}^{i_1}  \land   P_{2,1}^{i_2}  \land \dotsc   \land  P_{n,1}^{i_n}  \land  P_{n+1,1}^1 \dotsc  \land  P_{T,1}^1$}, where $Q_1^0$ means  that $M$ is in the initial state $q_0$, $S_{1,1}$ means  that the tape head scans square number 1,  $P_{n+1,1}^1 \dotsc  \land  P_{T,1}^1$ refers to  the blank symbols $\sigma_1$  after  $\sigma_{i_1}, \dotsc, \sigma_{i_n}$, and $P_{1,1}^{i_1}  \land   P_{2,1}^{i_2}  \land \dotsc   \land  P_{n,1}^{i_n}$  refers to  a string $\sigma_{i_1}, \dotsc, \sigma_{i_n}$ on the tape that  is claimed to refer to an instance $w$ of a problem. 

That is to say, given an instance $w$ of an $NP$ problem, a deterministic Turing machine $M$   determines whether   to accept  $w$  in polynomial time, and  $A(w)$ is built to represent this computation (see Fig. 1(b)). \\

In other words,   the proof of \textbf{Theorem 1}  claims that   a deterministic Turing machine can  accept $w$ like a nondeterministic Turing machine, that is, $NDTM$ is confused with    $DTM$ by means of  \textit{query machine}!

 \section{Interpretation of Cook's theorem based on  \textit{NDTM model}  }

Later researchers must have noticed   something  wrong  in the above proof, since  they completely abandoned   the  concept of query machine,  and proposed a \textit{NDTM  (NonDeterministic Turing Machine) model}   (see \cite{garey}, p. 30):

\textit{The NDTM model we will be using has exactly the same structure as a DTM  (Deterministic Turing Machine), except that it is augmented with a guessing module having its own write-only head.} \\

A computation of such a machine   takes place in two distinct stages (see  \cite{garey}, p. 30-31):

\textit{The first stage is the "guessing" stage. Initially, the input string $x$ is written in tape squares 1 through $\mid x \mid $ (while all other squares are blank), the read-write head is scanning square 1, the the write-only head is scanning square -1, and the finite state control is "inactive". The guessing module then directs the write-only head, one step at a time, either to write some symbol from $\Gamma$ in the tape square being scanned and move one square to left, or to stop, at which point the guessing module becomes inactive and the finite state control is activated in state $q_0$. The choice of whether to remain active, and, if so, which symbol from $\Gamma$ to write, is made by the guessing module in a totally arbitrary manner. Thus the guessing module can write any string from $\Gamma*$ before it halts and, indeed, need never halt.}

\textit{The "checking" stage begins when the finite state control is activated in state $q_0$. From this point on, the computation proceeds solely under the direction of the NDTM program according to exactly the same rules as for a DTM. The guessing module and its write-only head are no longer involved, having fulfilled their role by writing the guessed string on the tape. Of course, the guessed string can (and usually will) be examined during the checking stage. The computation ceases when and if the finite state control enters one of the two halt states (either $q_Y$ or $q_N$) and is said to be an accepting computation if it halts in state $q_Y$. All other computations, halting or not, are classed together simply as non-accepting computations. } \\

 \begin{figure} [h]
\begin{center}
\includegraphics[scale=.5]{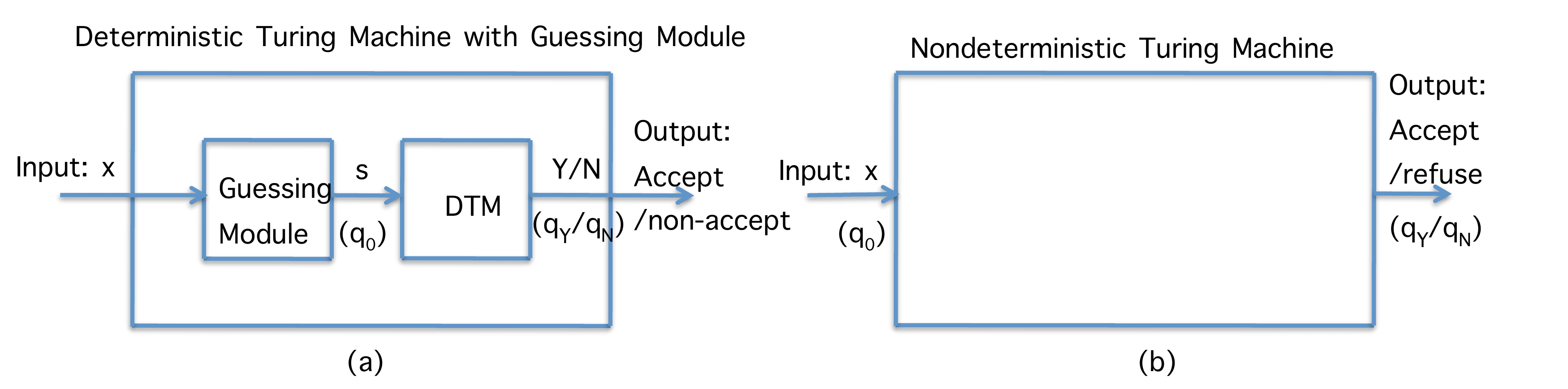}
\end{center}
\caption{(a) A machine of the NDTM model accepts $x$; (b) A nondeterministic Turing machine accepts $x$}
\label{fig5}
\end{figure}

That is, for a given instance $x$ of an $NP$ problem,  a guessing module finds a certificate $s$ of solution, then $s$ is  checked by a  deterministic Turing machine.  If  $s$ is a solution, the computation halts in state $q_Y$ and it is said to be an accepting computation; if $s$ is not a solution, the computation halts in state $q_N$ and it is said to be a  \textit{non-accepting computation}  (see Fig. 2(a)). 

However,  this   \textit{non-accepting computation}    is completely different from  the   \textit{refusing computation}   of a nondeterministic Turing machine  (see Fig. 2(b)), because this non-accepting computation has no sense, that is, a deterministic Turing machine with a guessing module  cannot determine whether $w$ is accepted or  refused in this case. In other words, the  \textit{NDTM model}    is not  $NDTM$. 

Unfortunately, it seems that these researchers did not really realize the  fundamental difference between the  \textit{NDTM model}  and  $NDTM$, and still attempted to construct a formula $A(w)$ in $CNF$ to represent a computation of a nondeterministic Turing machine through a machine of the $NDTM$ model. Consequently, the cognitive bases in Cook's theorem  have been hidden more deeply in terms of  the equivalence of  the \textit{solvability-based definition} and the \textit{verifiability-based one} (see \cite{sipser} section 7.3). 
   
    \section{Rectification of Cook's theorem}

  In fact, a formula $A(w)$ in $CNF$ can only represent a computation of a deterministic Turing machine, specially a  \textit{polynomial time verification}  for any problem in  $NP$ as well as  in $P$ (see Fig. 3).   \\
  
 \begin{figure} [h]
\begin{center}
\includegraphics[scale=.7]{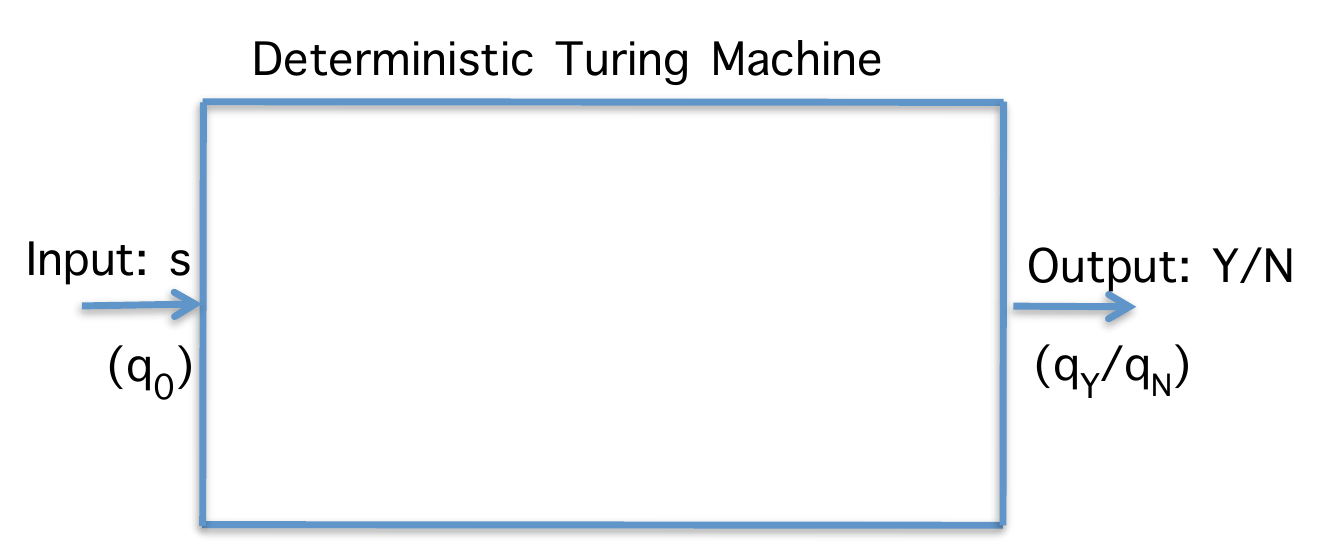}
\end{center}
\caption{A deterministic Turing machine verifies a certificate $s$ of solution}
\label{fig6}
\end{figure}

Let us  rectify  the  proof  of  \textbf{Theorem 1}.

At the initial time $t=1$,   a certificate $s$ of solution for an instance $w$ is put on the tape squares, then $E$  refers to  the assignment of the proposition symbols concerning the initial time $t=1$. From $E$,  the assignment of the proposition symbols concerning other times are deduced out according to $B  \land C   \land  D    \land  F   \land  G  \land  H$, and finally  the truth of  $I$ is deduced out. If $I=1$, it means that $s$ is a solution  for $w$; otherwise if $I=0$,  $s$ is not a solution for $w$.

Therefore,  $E$  consists of two parts : $E = E_1  \land E_2$, where $E_1=  Q_1^0   \land   S_{1,1} $ and $E_2 = P_{1,1}^{i_1}  \land   P_{2,1}^{i_2}  \land \dotsc   \land  P_{n,1}^{i_n}  \land  P_{n+1,1}^1 \dotsc  \land  P_{T,1}^1$. $E_1$ refers to  the initial state $q_0$  of $M$ as well as the tape head scanning square number 1,   while $E_2$    refers to a given certificate $s$. \\

Here, it should pay special attention to $E_2$. Since $E_2$ represents the assignment of  the proposition symbols concerning $s$ and $E_2=1$,  then $E_2$ should not appear in $A(w)$. Unfortunately, this key point has never been discussed in the literature  \cite{cook1}\cite{garey}, so that it   prevents from   interpreting correctly  the   meaning of $A(w)$.

Therefore, $B  \land C   \land  D  \land E_1   \land  F   \land  G  \land  H   \land  I$ refers to the verification about the truth of any certificate,   not limited to a given certificate $s$. Just in this sense, $B  \land C   \land  D  \land E_1   \land  F   \land  G  \land  H   \land  I$  becomes a function of $w$,  and  it can be denoted as   $A(w) = B  \land C   \land  D  \land E_1   \land  F   \land  G  \land  H   \land  I$. 

In other words, it is $B  \land C   \land  D  \land E_1   \land  F   \land  G  \land  H   \land  I$, rather than $B  \land C   \land  D  \land E   \land  F   \land  G  \land  H   \land  I$,    that is   intended  to represent an instance $w$ of a problem in terms of $A(w)$. \\

Now,  we can clarify the real meaning of $A(w)$, denoted in \textbf{Theorem 0}: 

\textbf{Theorem 0}
\textit{Any problem  verifiable by a deterministic Turing machine in polynomial time  can be represented as    $A(w)$.} \\

Since an $NP$ problem   is  polynomially verifiable according to the definition of $NDTM$,  so an $NP$ problem  can be represented as  $A(w)$.  Note that  any $P$ problem is   polynomially verifiable,  so it  can be  also represented as  $A(w)$ in the same way.  \\

Furthermore, determining the satisfiability of $A(w)$ corresponds to determining   the existence of solution for  an $NP$ problem, thus   it  deduces out  \textbf{Theorem 1'} in terms of  the usual expression  of Cook's theorem:

\textbf{Theorem 1'}
\textit{Any problem solvable  by a nondeterministic Turing machine in polynomial time can be represented as a $SAT$ problem.} \\

\textbf{Theorem 1'} consists of   the rectification of the original statement \textbf{Theorem 1} of Cook's theorem including its proof. Therefore,  \textbf{Theorem 1} is just a corollary of \textbf{Theorem 0}, and  the relation between  \textbf{Theorem 0} and \textbf{Theorem 1} is the  cause-effect relationship. In other words, the two definitions of $NP$,  the   verifiability-based definition corresponding to \textbf{Theorem 0} and the solvability-based  one to  \textbf{Theorem 1}, are not equivalent, because they  are situated at different levels of concept and have the  cause-effect relationship.

\section{Conclusion}

In this paper, we give preliminary interpretation of Cook's theorem in \cite{cook1} and reveal the cognitive biases in the proof of Cook's theorem,  which leads to  the  confusion between different levels of     \textit{nondeterminism} and  \textit{determinism}, thus  the confusion of   the solvability-based definition and the verifiability-based one  of $NP$, finally causes the loss of  \textit{nondeterminism}   from   $NP$.  

The work   argues again that the difficulty in understanding   \textit{P versus NP}   lies at  firstly  cognition level, then logic  level  \cite{zhou2}.  A similar opinion is also suggested in \cite{scott}.

We hope that this work can evoke reflections from different angles about some fundamental problems in cognitive science, and contribute to  the understanding of   \textit{P  versus NP}. Furthermore, we hope that this work can help to understand the complementarity of Chinese thought and Western philosophy.


\end{document}